# Effects of the Generalized Uncertainty Principle on Quantum Tunneling

**Gardo Blado, Trevor Prescott, James Jennings, Joshuah Ceyanes and Rafael Sepulveda**

*College of Science and Mathematics*

*Houston Baptist University*

*7502 Fondren Rd., Houston, Texas, U.S.A*

**Abstract**

In a previous paper[1], we showed that quantum gravity effects can be discussed with only a background in non-relativistic quantum mechanics at the undergraduate level by looking at the effect of the generalized uncertainty principle (GUP) on the finite and infinite square wells. In this paper, we derive the GUP corrections to the tunneling probability of simple quantum mechanical systems which are accessible to undergraduates (alpha decay, simple models of quantum cosmogenesis and gravitational tunneling radiation) and which employ the WKB approximation, a topic discussed in undergraduate quantum mechanics classes. It is shown that the GUP correction increases the tunneling probability in each of the examples discussed.

## I. Introduction

One of the most exciting problems in theoretical physics is the problem of the formulation of quantum gravity which attempts to unite the two pillars of modern physics namely general relativity and quantum mechanics. The vast literature of quantum gravity effects require a background in general relativity and quantum field theory. However the study of quantum gravity effects via the modification of the Heisenberg uncertainty principle to a generalized uncertainty principle due to quantum gravity theories like loop quantum gravity and string theory, is very accessible to undergraduates with a background in quantum mechanics. In a previous paper[1], we discussed how GUP corrections can be calculated using the common square well potentials in undergraduate quantum mechanics. In the present paper we continue to explore topics which are understandable to undergraduates, namely tunneling effects using the WKB approximation as applied to simple models.

The time independent Schrodinger equation (TISE) $-\frac{\hbar^2}{2m}\frac{d^2\psi}{dx^2} + V(x)\psi(x) = E\psi(x)$ can be written as $\widehat{H}\psi = E\psi$ with the Hamiltonian operator

$$\widehat{H} = \frac{\widehat{p}^2}{2m} + V \quad (1)$$

and momentum operator $\widehat{p} = \frac{\hbar}{i}\frac{d}{dx}$. The momentum position operators satisfy a commutation relation,

$$[\widehat{x},\widehat{p}] = i\hbar. \quad (2)$$

The general uncertainty inequality for any two observables $A$ and $B$ given by[2] $(\Delta A)^2(\Delta B)^2 \geq \left(\frac{1}{2i}\langle[\widehat{A},\widehat{B}]\rangle\right)^2$ with $\widehat{A} = \widehat{x}$ and $\widehat{B} = \widehat{p}$ gives rise to the uncertainty relation $\Delta x \Delta p \geq \frac{\hbar}{2}$ or $\Delta x \geq \frac{\hbar}{2\Delta p}$. Quantum gravity theories (such as string theory) modifies equation (2) as[3]

$$[x,p] = i\hbar[1 + \beta p^2] \quad (3)$$

which gives rise to the generalized uncertainty principle (GUP)[3]

$$\Delta x \Delta p \geq \frac{\hbar}{2}[1 + \beta(\Delta p)^2 + \beta\langle p\rangle^2]$$

with $\beta$ as the GUP correction parameter which is small. It can be shown that equation (3) can be satisfied if we set in coordinate space[4]

$$x = x_0, p = p_0(1 + \beta p_0^2) \quad (4)$$

where $x_0$ and $p_0$ satisfy the usual commutation relation $[x_0, p_0] = i\hbar$.

The modification of the momentum operator in equation (4) modifies the Hamiltonian in equation (1) which in turn modifies the Schrodinger equation.

## II.  WKB-GUP

The WKB approximation is a typical topic in undergraduate quantum mechanics classes (chapter 8 of reference [2] for example). Given the TISE $-\frac{\hbar^2}{2m}\frac{d^2\psi}{dx^2} + V(x)\psi(x) = E\psi(x)$ with a non-constant but slowly varying potential $V(x)$, the WKB approximation gives the wavefunction as $\psi(x) \cong \frac{C}{\sqrt{p(x)}} e^{\pm \frac{i}{\hbar}\int p(x)\,dx}$ where $p(x) = \sqrt{2m(E - V(x))}$ with $E > V(x)$ and, $\psi(x) \cong \frac{C}{\sqrt{|p(x)|}} e^{\pm \frac{1}{\hbar}\int |p(x)|\,dx}$ where

$$|p(x)| = \sqrt{2m(V(x) - E)} \qquad (5)$$

with $V(x) > E$. The tunneling probability can be calculated as

$$T \cong e^{-2\gamma} \qquad (6)$$

with

$$\gamma = \frac{1}{\hbar}\int |p(x)|\,dx \qquad (7)$$

in the case of a broad and high barrier (Problem 8.10 of reference [2]).

The GUP-modified wave function based on equation (3), can be derived as[5] (for $E > V(x)$)

$$\psi(x) = \frac{1}{\sqrt{p(1+\beta p^2)}}\left(Ae^{\frac{i}{\hbar}\int p\,dx} + Be^{-\frac{i}{\hbar}\int p\,dx}\right) \qquad (8)$$

with $p(x) = \frac{1}{\sqrt{\beta}}\arctan(\sqrt{\beta}[\sqrt{2m(E-V(x))}])$.

Clearly, for $V(x) > E$, we can write a similar equation,

$$\psi(x) = \frac{1}{\sqrt{p(1+\beta p^2)}}\left(Ce^{\frac{1}{\hbar}\int |p|\,dx} + De^{-\frac{1}{\hbar}\int |p|\,dx}\right) \qquad (9)$$

where

$$|p(x)| = \frac{1}{\sqrt{\beta}}\arctan\left(\sqrt{\beta}\left[\sqrt{2m(V(x)-E)}\right]\right) \qquad (10)$$

and $p(x)$ and $|p(x)|$ above reduce to their corresponding non-GUP expressions in the limit when $\beta \to 0$ with the $\beta$ in equations (3) and (4).

Using equation (8) to equation (10), one can show, following the method outlined in Problem 8.10 of reference [2] and the GUP correction discussion in reference [6] (or by inspection similar to the discussion of reference [7]) that the GUP-modified tunneling probability $T_{GUP}$ is given by

$$\begin{cases} a) T_{GUP} \cong e^{-2\gamma_{GUP}} \\ b) \gamma_{GUP} = \frac{1}{\hbar} \int |p(x)| \, dx \\ c) |p(x)| = \frac{1}{\sqrt{\beta}} \arctan\left(\sqrt{\beta}\left[\sqrt{2m(V(x)-E)}\right]\right) \end{cases} \quad (11)$$

### III. Alpha Decay

Let us first discuss a familiar system which is typically discussed in an undergraduate quantum mechanics class which illustrates the application of the WKB method, namely the alpha decay (Ch 8 of reference [2]). The potential is illustrated in Figure 1 in which $r_1$ is the radius of the nucleus and $r_2$ is the distance at which the energy of the alpha particle (with charge $2e$) equals the Coulomb potential due to the nucleus (with charge $Ze$ due to $Z$ protons) given by

$$V(r) = \frac{Ze(2e)}{4\pi\epsilon_0 r} = \frac{2Ze^2}{4\pi\epsilon_0 r} \quad (12)$$

At $r_2$, the energy $E$ is equal to the potential energy. The Schrodinger equation is given by

$-\frac{\hbar^2}{2m}\frac{d^2\psi}{dx^2} + \frac{2Ze^2}{4\pi\epsilon_0 r} \psi(r) = E\psi(r)$ with

$$E = \frac{2Ze^2}{4\pi\epsilon_0 r_2} \quad (13)$$

Hence

$\gamma = \frac{1}{\hbar}\int_{r_1}^{r_2} \sqrt{2m(V(r)-E)} \, dr = \frac{1}{\hbar}\int_{r_1}^{r_2} \sqrt{2m\left(\frac{2Ze^2}{4\pi\epsilon_0 r} - \frac{2Ze^2}{4\pi\epsilon_0 r_2}\right)} \, dr$. Using equation (13), we can rewrite the equation as

$$\begin{aligned} \gamma &= \frac{\sqrt{2mE}}{\hbar} \int_{r_1}^{r_2} dr \sqrt{\frac{r_2}{r}-1} \\ &= \frac{\sqrt{2mE}}{\hbar}\left(-\frac{1}{2}r_2 \sin^{-1}\left(\frac{2r_1-r_2}{r_2}\right) + \frac{\pi r_2}{4} - \sqrt{r_1(r_2-r_1)}\right) \end{aligned} \quad (14)$$

which can be shown to agree with reference [2] by using trigonometric identities. The tunneling probability is then given by equation (6) with $\gamma$ given by equation (14).

To find the GUP correction, we use the modified tunneling probability in equation (11).

$$\gamma_{GUP} = \frac{1}{\hbar}\int |p(r)|\, dr = \frac{1}{\hbar}\int_{r_1}^{r_2} \frac{1}{\sqrt{\beta}} \arctan\left(\sqrt{\beta}\left[\sqrt{2m(V(r)-E)}\right]\right) dr$$

with the potential given by equation (12). To facilitate the calculation, we calculate up to first order in $\beta$ which yields (with $\tan^{-1} x \cong x - x^3/3$),

$$\gamma_{GUP} \cong \gamma + \Delta\gamma_{GUP}$$

with

$\Delta\gamma_{GUP} = -\frac{(2mE)^{3/2}}{3\hbar}\beta \int_{r_1}^{r_2}\left(\frac{r_2}{r}-1\right)^{3/2} dr$ which leads to $\Delta\gamma_{GUP} = \frac{(2mE)^{3/2}}{12\hbar}\beta\, g(r_2)$ where

$$g(r_2) = g(E) = -6r_2 \sin^{-1}\left(\frac{2r_1 - r_2}{r_2}\right) + 3\pi r_2 + \frac{4r_1^{3/2}}{\sqrt{r_2 - r_1}} + \frac{4r_2 r_1^{1/2}}{\sqrt{r_2 - r_1}} \tag{15}$$

$$-\frac{8r_2^2}{\sqrt{r_1(r_2 - r_1)}}$$

with $r_2$ given in terms of the energy $E$ in equation (13). We can show that $\Delta\gamma_{GUP} < 0$. For uranium with nuclear radius $r_1 = 9.3 \times 10^{-15}$ m, we plot $g(E)$ vs. the energy (for $E = 0$ to $44.8 \times 10^{-13}$ J or 28 Mev) in Figure 2 which shows that $g(E) < 0$. Hence $\Delta\gamma_{GUP} < 0$ and

$$T_{GUP} \cong e^{-2\gamma_{GUP}} = e^{-2\gamma} e^{-\Delta\gamma_{GUP}} = T e^{-\Delta\gamma_{GUP}} > T$$

where $T$ is the tunneling probability $e^{-2\gamma}$ with $\gamma$ given by equation (14). The GUP correction increases the tunneling probability which is consistent with the result of reference [7] where the authors used non-commutative geometry. Just as in reference [7], the GUP correction is too small to be of interest.

## IV. Quantum Cosmogenesis

In this section, we discuss a very simple model of quantum cosmogenesis[8] that is very accessible to undergraduates. We consider an isotropic, homogeneous universe described by the Friedmann-Robertson-Walker (FRW) line element given by $ds^2 = -c^2 dt^2 + a^2(t)\left(\frac{dr^2}{1-kr^2} + r^2[d\theta^2 + \sin^2\theta\, d\phi^2]\right)$ where $a(t)$ is the cosmic scale factor which measures the universe's size. The time evolution of the cosmic scale factor is described by the Einstein equation

$$\dot{a}^2 + \left(1 - \frac{\Lambda}{3}a^2\right) = 0 \tag{16}$$

where the cosmological term $\Lambda$ is given by

$$\Lambda = 8\pi G \rho_{\text{vac}} \tag{17}$$

with $G$ as Newton's constant and $\rho_{\text{vac}}$ is the constant vacuum energy density. One can apply canonical quantization to equation (16) to get the Wheeler-Dewitt equation (WDE)

$$\left[\frac{d^2}{da^2} - \left(\frac{3\pi}{2G}\right)^2 a^2\left(1 - \frac{a^2}{a_0^2}\right)\right]\psi(a) = 0 \tag{18}$$

where

$$a_0 = \sqrt{\frac{3}{\Lambda}} \tag{19}$$

and $\psi(a)$ is the wave function of the universe. Note that the WDE looks like the TISE. Here, we use Planck units ($\hbar = 1$) and one-half unit mass[8]. Applying the WKB approximation, one can derive the tunneling probability that the universe pops into existence from zero size ($a = 0$) and zero energy to a finite size $a = a_0$, given by $T \cong e^{-2\gamma}$ with (as before)

$$\gamma = \frac{1}{\hbar}\int |p(a)|\, da = \frac{1}{\hbar}\int_0^{a_0} da\, \sqrt{2m(V(a) - E)} \tag{20}$$

The potential can be gleaned from the WDE, equation (18), to be

$$V(a) = \left(\frac{3\pi}{2G}\right)^2 a^2\left(1 - \frac{a^2}{a_0^2}\right) \tag{21}$$

With $E = 0$, it is shown that[8]

$$T = e^{-3/(8G^2\rho_{vac})} \tag{22}$$

To find the GUP correction, we use the modified tunneling probability in equation (11). We calculate

$$\gamma_{GUP} = \frac{1}{\hbar}\int da|p(a)| = \int_0^{a_0} da\, \frac{1}{\sqrt{\beta}} \arctan\left(\sqrt{\beta}\left[\sqrt{2m(V(a) - E)}\right]\right)$$

with zero energy and the potential $V(a)$ in equation (21). To facilitate the calculation, we calculate up to first order in $\beta$ which yields (with $\tan^{-1} x \cong x - x^3/3$),

$$\gamma_{GUP} \cong \gamma - \beta\frac{(2m)^{3/2}}{3\hbar}\cdot\left(\frac{3\pi}{2G}\right)^3 \int_0^{a_0} da\, a^3\left(1 - \frac{a^2}{a_0^2}\right)^{3/2} \tag{23}$$

where $\gamma$ is the usual term in equation (20) without the GUP correction. The integral can be computed readily to yield $\frac{2}{35}a_0^4$. With equation (17) and equation (19), we rewrite $a_0$ in terms of the constant vacuum energy density $\rho_{\text{vac}}$. Equation (23) becomes (with Planck units ($\hbar = 1$) and one-half unit mass)

$$\gamma_{GUP} \cong \gamma - \beta \frac{9\pi}{140\, G} \left( \frac{3}{8G^2 \rho_{vac}} \right)^2$$

which yields from equation (11),

$$\begin{cases} a)\ T_{GUP} \cong e^{-2\gamma_{GUP}} = e^{-2\gamma}\, e^{\beta \delta} = T e^{\beta \delta} \\ b)\ \delta = \frac{9\pi}{70\, G} \left( \frac{3}{8G^2 \rho_{vac}} \right)^2 \end{cases} \quad (24)$$

where $T$ is the tunneling probability of equation (22). Note that the GUP correction increases the tunneling probability when we consider the effect of the presence of minimal length. This is expected from the discussion in reference [9] in which quantum gravity effects become large at small scales. Let us compute a numerical estimate of $T_{GUP}/T = e^{\beta \delta}$ in equation (24). As in reference [8], we set $a_0^2 = G$ and using equation (17), equation (19), and in Planck units, we get $\delta = \frac{9\pi^3}{70}$. We set $\beta = 0.01$[1] and consequently get $T_{GUP}/T = e^{\beta \delta} \approx 1.04$. Hence, in this simple case, the GUP correction can increase the tunneling probability by 4%.

## V. Gravitational Tunneling Radiation

We now turn to another simple model discussed by Rabinowitz[10]. In his paper, he proposed a possible explanation on why the Hawking radiation has yet to be observed. His main result is the derivation of the relationship between the radiated power from a black hole by gravitational radiation due to tunneling, $P_R$ and the radiated power due to Hawking radiation[11,12], $P_{SH}$ given by

$$P_{SH} \approx \frac{P_R}{e^{-2\gamma}} = \frac{P_R}{T} \quad (25)$$

where $T$ is the tunneling probability of gravitational radiation similar to equation (6). In the presence of another body in the vicinity of the black hole, the tunneling probability $T$ is shown to increase thereby decreasing the radiated power due to Hawking radiation, $P_{SH}$. We give more details in the following discussion.

As a first approximation, reference [10] considered a black hole centered at the origin with mass $M$, radius $R_H$ and a second body with mass $M_2$ centered at $R_2$. To calculate the tunneling probability for a particle of mass *m* using equation (5) to equation (7), he considers the following one-dimensional Schrodinger equation.

$$-\frac{\hbar^2}{2m}\frac{d^2\psi}{dr^2} + \frac{-GmM_2}{R_2 - r}\psi = E\psi \quad (26)$$

The potential is shown in Figure 3. The black hole is represented by a square well. The presence of the second body centered at $R_2$ essentially lowers the barrier giving it a finite width which then allows the tunneling. Note that from Figure 3, with $r = \beta_2$ as a turning point, we can write the energy as

$$E = \frac{-GmM_2}{R_2 - \beta_2} \quad (27)$$

From equations (7), (5), the potential in equation (26) and equation (27), we get

$$\gamma = \frac{1}{\hbar}\int_{R_H}^{\beta_2}\sqrt{2m(V(r)-E)}\,dr = \frac{1}{\hbar}\int_{R_H}^{\beta_2}\sqrt{2m\left(\frac{-GmM_2}{R_2-r} - \frac{-GmM_2}{R_2-\beta_2}\right)}\,dr$$

which yields[10],

$$\gamma = \frac{m}{\hbar}\sqrt{2GM_2}\left\{\sqrt{\frac{(R_2-R_H)(\beta_2-R_H)}{R_2-\beta_2}} - \sqrt{R_2-\beta_2}\ln\left(\frac{\sqrt{R_2-R_H}+\sqrt{\beta_2-R_H}}{\sqrt{R_2-\beta_2}}\right)\right\}.$$

The GUP correction can be calculated by using equation (11), as before with the potential given in equation (26) and the energy in equation (27). Calculating only to first order in $\beta$, we get $\gamma_{GUP} = \gamma + \Delta\gamma_{GUP}$ with

$$\Delta\gamma_{GUP} = -\beta\frac{\left(m\sqrt{2GM_2}\right)^3}{3\hbar}\left\{-\frac{3}{\sqrt{R_2-\beta_2}}\ln\left(\frac{\sqrt{R_2-R_H}+\sqrt{\beta_2-R_H}}{\sqrt{R_2-\beta_2}}\right) \right. \quad (28)$$
$$\left. + \frac{\sqrt{(R_2-R_H)(\beta_2-R_H)}}{(R_2-\beta_2)^{3/2}}\left(1+\frac{2(R_2-\beta_2)}{R_2-R_H}\right)\right\}$$

It can be shown that $\Delta\gamma_{GUP} < 0$. We let $\beta_2 = k_1 R_H$ and $R_2 = k_2 R_H$ with $1 < k_1 < k_2$ in equation (28) to get

$$\Delta\gamma_{GUP} = -\beta\frac{\left(m\sqrt{2GM_2}\right)^3}{3\hbar\sqrt{R_H}}F(k_1, k_2)$$

with

$$F(k_1, k_2) = -\frac{3}{\sqrt{k_2 - k_1}} \ln\left[\frac{\sqrt{k_2 - 1} + \sqrt{k_1 - 1}}{\sqrt{k_2 - k_1}}\right] + \frac{\sqrt{(k_2 - 1)(k_1 - 1)}}{(k_2 - k_1)^{3/2}}\left[1 + \frac{2(k_2 - k_1)}{k_2 - 1}\right] \quad (29)$$

We graph $F(k_1, k_2)$ in Figure 4 where we see it to be positive which makes $\Delta\gamma_{GUP} < 0$. Recalling that $T_{GUP} \cong e^{-2\gamma_{GUP}} = e^{-2(\gamma + \Delta\gamma_{GUP})} = e^{-2\gamma} e^{-2\Delta\gamma_{GUP}} = T e^{-2\Delta\gamma_{GUP}}$ and with $\Delta\gamma_{GUP} < 0$, we see that the tunneling probability is increased in the presence of the GUP correction which further suppresses the $P_{SH}$ in equation (25).

## VI. Conclusions

Using the alpha decay and simple models of quantum cosmogenesis and black holes, we are able to explore quantum gravity effects through the generalized uncertainty principle using concepts studied in undergraduate physics. In each of the cases we studied, we showed that there is an increase in the tunneling probabilities when the GUP correction is considered. We claim without proof that this might be generally the case. In fact, using the simple model of reference [8], we showed that the probability that the universe pops into existence from zero size ($a = 0$) and zero energy to a finite size $a = a_0$ increases by about 4%.

Having an expression for the GUP-corrected tunneling probability in equation (11), we can further investigate the different scenarios discussed in reference [10] namely the case in which both the gravitational potentials of the black hole and a nearby body are given by $V(r) \sim 1/r$ instead of approximating the potential of the black hole to be a square well as in Figure 3 and the discussion of the Aharonov-Bohm-like effect in which the black hole is surrounded by a spherical shell. It will also be interesting to show generally that the GUP-corrected tunneling probability given by equation (11) increases the uncorrected tunneling probability.

## VII. Acknowledgements

The authors would like to thank Dr. James Claycomb of the Houston Baptist University Physics Department for suggesting the alpha decay and quantum cosmogenesis problems.

# Figures:

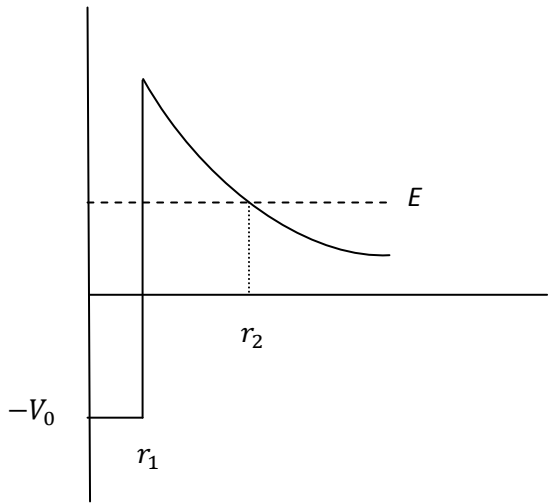

**Figure 1:** Potential for alpha decay[2]

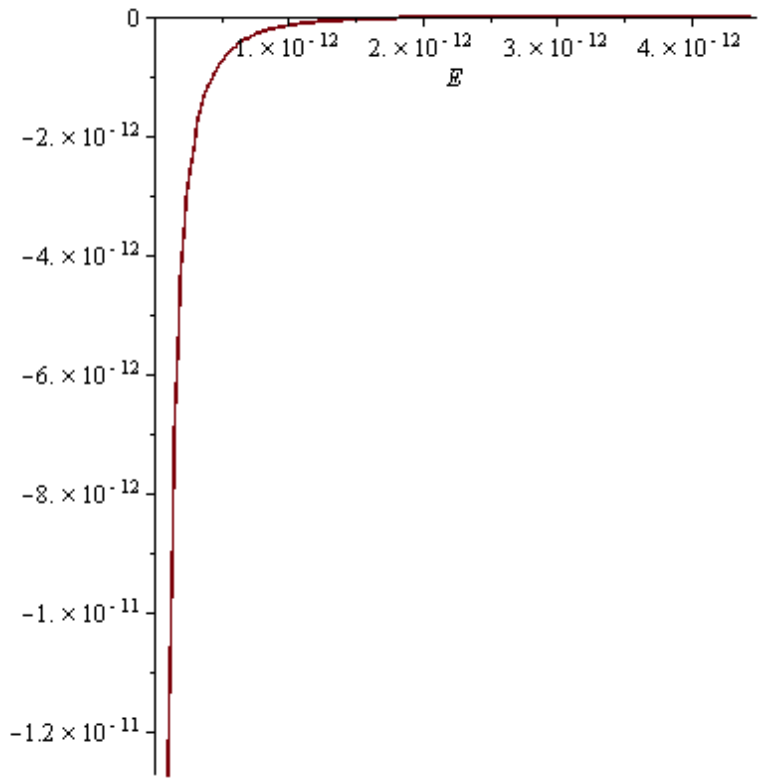

**Figure 2:** Plot of $g(E)$ vs. the energy $E$ of equation (15).

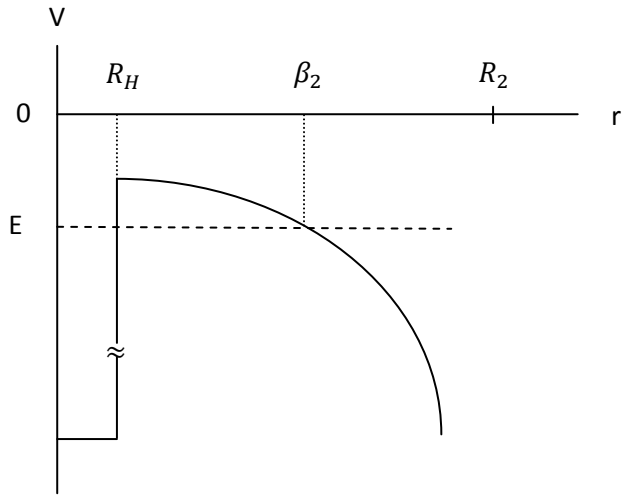

**Figure 3:** Gravitational potential energy of mass $M_2$ at $R_2$ with a black hole of radius $R_H$ at the origin[10].

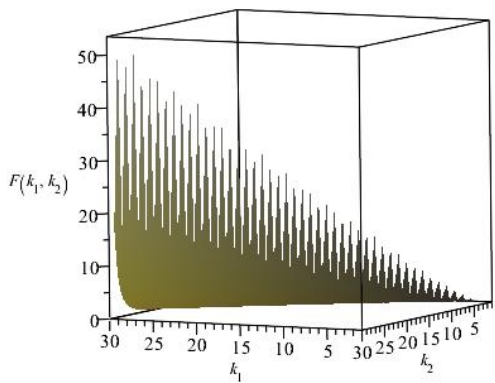

**Figure 4:** Graph of $F(k_1, k_2)$ of equation (29).

# REFERENCES:

[1] Blado G, Owens C, and Meyers V 2014 Quantum Wells and the Generalized Uncertainty Principle *Eur. J. Phys.* **35** 065011

[2] Griffiths D J 2005 *Introduction to Quantum Mechanics* (New Jersey: Prentice Hall)

[3] Kempf A, Mangano G and Mann R B 1995 Hilbert space representation of the minimal length uncertainty relation *Phys. Rev. D* **52** 1108-18

[4] Das S and Vagenas E C 2009 Phenomenological Implications of the Generalized Uncertainty Principle *Can. J. Phys.* **87** 233-40

[5] Fityo T V, Vakarchuk I O and Tkachuk V M 2008 WKB approximation in deformed space with the minimal length and minimal momentum *Journal of Physics A: Mathematical and Theoretical* **41** 045305

[6] Tao J, Wang P and Yang H 2012 Homogeneous Field and WKB Approximation In Deformed Quantum Mechanics with Minimal Length *arxiv* 1211.5650

[7] Sprenger M, Nicolini P and Bleicher M 2012 An Introduction to Minimal Length Phenomenology *Eur. J. Phys* **33** 853-862

[8] Atkatz D 1994 Quantum Cosmology for Pedestrians *Am. J. Phys* **62** 619-627

[9] Linde A D 1984 Quantum Creation of the Inflationary Universe *Lettere Al Nuovo Cimento* **39** 401-405

[10] Rabinowitz M 1999 Gravitational Tunneling Radiation *Phys Essays*, **12,** 346-357

[11] Hawking S W 1974 Black hole explosions? *Nature* **248** 30-31

[12] Hawking S W 1975 Particle Creation by Black Holes *Commun. Math. Phys.* **43** 199-220